# Formation of Curved Micron-Sized Single Crystals

*Maria Koifman Khristosov*[1,2], *Lee Kabalah-Amitai*[1], *Manfred Burghammer*[3,4], *Alex Katsman*[1] *and Boaz Pokroy*[1,2]

(1)Department of Materials Science and Engineering, Technion Israel Institute of Technology, 32000 Haifa, Israel.

(2)Russell Berrie Nanotechnology Institute, Technion Israel Institute of Technology, 32000 Haifa, Israel.

(3)European Synchrotron Radiation Facility, BP 220, F-38043 Grenoble Cedex, France.

(4)X-ray Microspectroscopy and Imaging Research Group, Department of Analytical Chemistry Ghent University, B-9000 Ghent, Belgium.

E-mail: bpokroy@technion.ac.il

ABSTRACT

Crystals in nature often demonstrate curved morphologies rather than classical faceted surfaces. Inspired by biogenic curved single crystals, we demonstrate that gold single crystals exhibiting curved surfaces can be grown with no need of any fabrication steps. These single crystals grow from the confined volume of a droplet of a eutectic composition melt which forms *via* the dewetting of nanometric thin films. We can control their curvature by controlling the environment in which the process is carried out, including several parameters, such as the contact angle and the curvature of the drops by changing the surface tension of the liquid drop during crystal growth. Here we present an energetic model that explains this phenomenon and predicts why and under what conditions crystals will be forced to grow with the curvature of the micro-droplet even though the energetic state of a curved single crystal is very high.

In nature, biogenic crystals are often formed *via* amorphous precursors rather than *via* classical nucleation and growth. One remarkable outcome of crystal growth *via* an amorphous precursor is the ability of the organism to obtain unfaceted sculptured single crystals with rounded shapes and even such crystals that are highly porous. Another outstanding advantage of growing crystals *via* this non-classical route is the fact that they can be molded into any desired shape.[1-4] Crystals grown by classical methods of nucleation and growth, have facets dictated by the atomic structure and minimization of the surface energy, when the facets are the low energy crystallographic planes.[5] Attempts have been made to mimic the growth of biogenic crystals with intricate shapes and morphologies.[6] Thus, for example, single crystals of calcite were grown through the amorphous calcium carbonate precursor phase on micropatterned templates induced by a self-assembled monolayer on the template surface;[7] or, as another example, using a sponge-like polymer membrane as a template, single crystals of calcium carbonate, lead(II)sulfate, strontium sulfate and others were grown while being constrained into intricate shapes replicating sea urchin skeletal plates.[8-11] These processes were carried out using ceramic materials found in biogenic material such as $CaCO_3$, but no such attempts have been made to date with widely used functional technological materials such as metals. Fashioning single crystals with curved shapes from functional materials such as noble metals would have high research and technological potential, for example in photonics (micro-lenses and micro-mirrors).[12, 13]

With such functional materials as the starting point, however, obtaining the curved morphologies achieved in biogenic crystals requires additional fabrication steps, such as sculpturing, drilling, and polishing the single crystals, or using standard nano- and micro-fabrication techniques.[12, 14, 15] To the best of our knowledge these procedures have not been applied on micron-sized single crystals. Polishing and microfabrication have been carried out on curved crystals obtained with gold, but the curvature was on a scale of millimeters and not microns.[14]

In the present study we were able to produce curved micron-sized gold single crystals by utilizing confined droplets within which a single growing crystal could be confined. The idea was that if we could control the size and shape of the droplet, we would be able to control the size and shape of the single crystal growing inside it. The mechanism we employed for the formation of such droplets was the dewetting phenomenon of nanometric thin films. To introduce dewetting into liquid droplets a thin film needs to be brought to its melting point. A eutectic system such as gold-germanium seemed an ideal choice for this purpose owing to relatively low temperature at which the liquid state can be achieved ($T_E = 361°C$).[16] If a eutectic film of Au-Ge is deposited on a non-wetting surface such as $SiO_2$, on reaching the eutectic temperature it dewets and produces liquid droplets.

Metal nano- and micro-nanoparticles are used widely in a variety of applications, such as magnetic memory arrays,[17, 18] plasmonic waveguides,[19] plasmonic sensors,[20] biological sensors[21] and catalysis of other structures such as nanowires and nanofiber growth.[22, 23] One of the applications for Au nanoparticles is catalytic growth of nanostructures such as carbon nanotubes[24] and germanium nanowires.[25] Growing of metal nano- and micro-particles can be achieved by different methods, for example by self-assembly chemical processes,[26] lithography[15] or dewetting of thin films through solid[27] or liquid state dewetting.[28] The dewetting process can be reached by different techniques, such as thermal annealing,[29] electron beam[28] or lasers.[14]

Here we produced Au-Ge thin films by evaporating thin layers of Au and Ge on oxidized Si wafers. The thicknesses of the films were altered (Au 100-200 nm, Ge 50-100 nm) according to the composition ratio needed. For example, to achieve a eutectic melt composition (28 at% Ge), the Au and Ge thin films were 150 nm and 78 nm thick, respectively. Samples were annealed at 550°C for durations varying between 10 min and 3 h, and then cooled to a room temperature at the rate of 200°Cmin$^{-1}$. The annealing environment was N$_2$ (99.999%), with or without pump purging prior to annealing (pump purging reached 10$^{-3}$ torr before annealing).

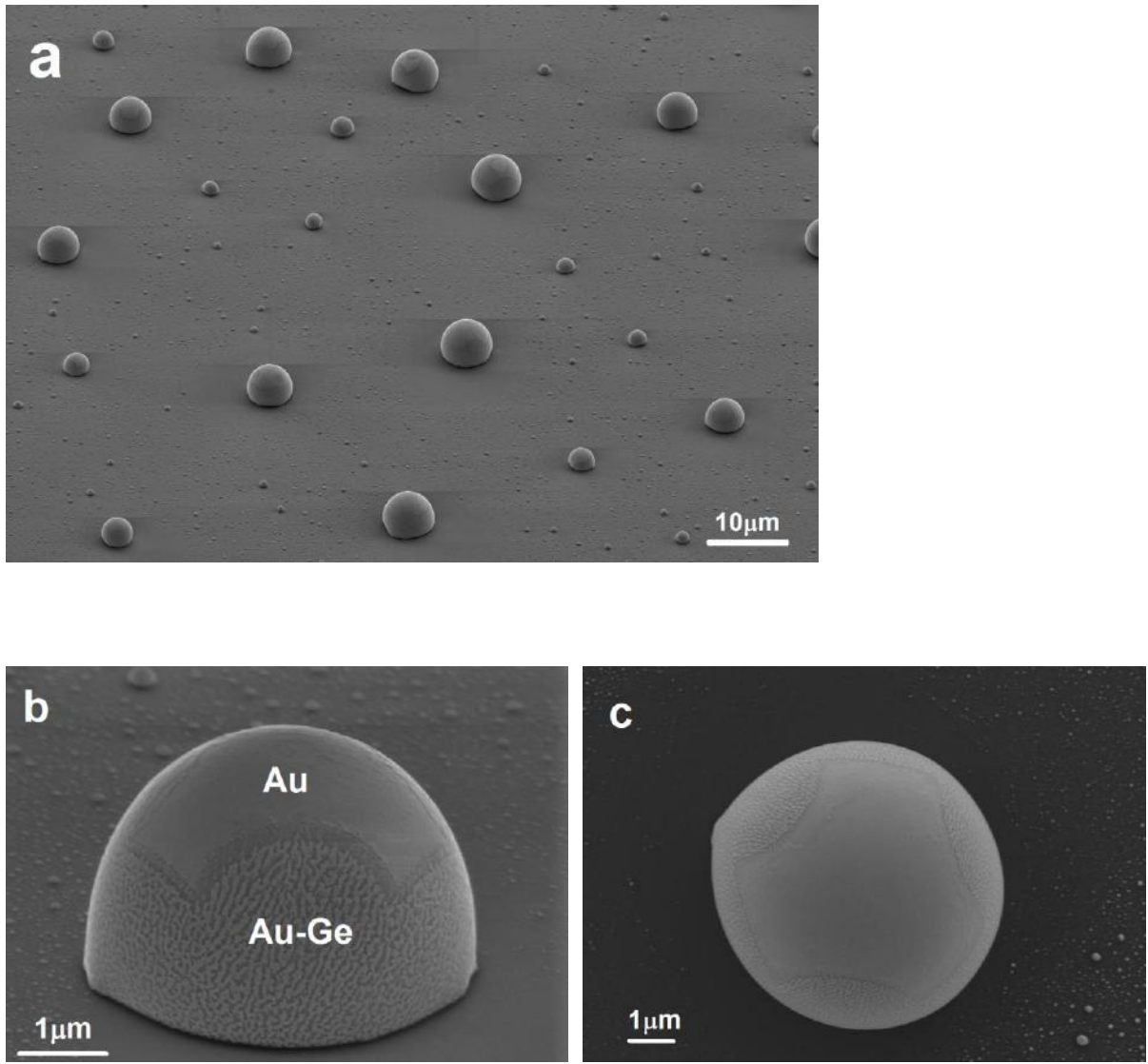

**Figure 1.** HRSEM micrographs of Au-Ge droplets after dewetting. (a) A large area view of the dewetted surface revealing several droplets (52° tilt). (b) A side view (60° tilt) of a curved gold single crystal within a micro droplet. (c) A top view of the droplet in (b).

During the annealing process, dewetting led to the formation of droplets of Au-Ge under which Au single crystals were formed (Figure 1). Energy-dispersive X-ray Spectroscopy (EDS) examination of the single crystals proved that they were gold and that the surrounding eutectic phase contained a combination of gold, germanium and oxygen (oxidized Ge) (Supporting Information Figure S1). The single crystals inside the droplets were further examined by Electron Backscatter Diffraction (EBSD) (Supporting Information Figure S2). The EBSD mapping of these gold crystals over an area of 5×5 μm$^2$ revealed identical electron backscatter diffraction patterns (EBSP) over the entire scanned area, affirming that these were indeed single crystals. As the EBSD signal comes from nearly 100 nm below the surface, we wanted to further confirm that these crystals are single throughout their entire volume. To this end we prepared cross-sections of these droplets utilizing a Focused Ion Beam (FIB). This enabled us to observe that each crystal was faceted inside the droplet but possessed a curved surface confined by the droplet's surface. Moreover, the 3D curvature of these single crystals was identical to that of the droplets (Figure 2(a)). Using FIB we sliced about a hundred droplets, all of which revealed similar internal crystal morphologies and curved surfaces.

In our opinion, the only way to create a curved surface from a single crystal of gold is by the formation of atomic steps on the surface, while a higher degree of curvature will lead to a higher density of atomic steps.[14] The morphology of the faceted area of the gold single crystals inside a droplet, is a truncated octahedron, which is the most preferable equilibrium shape of gold (Figure 2(c)).[30] The exposed facets within the droplet (not on the droplet's surface) reveal the {111} and {110} planes of the truncated octahedron. The facets that can be seen in the FIB cross sections depend on the relative orientation of the cutting during the FIB cross section preparation. In relation to the orientation of the single crystals of gold relatively to the substrate, we can conclude that in most cases the orientation is that of the [111] or close to it.

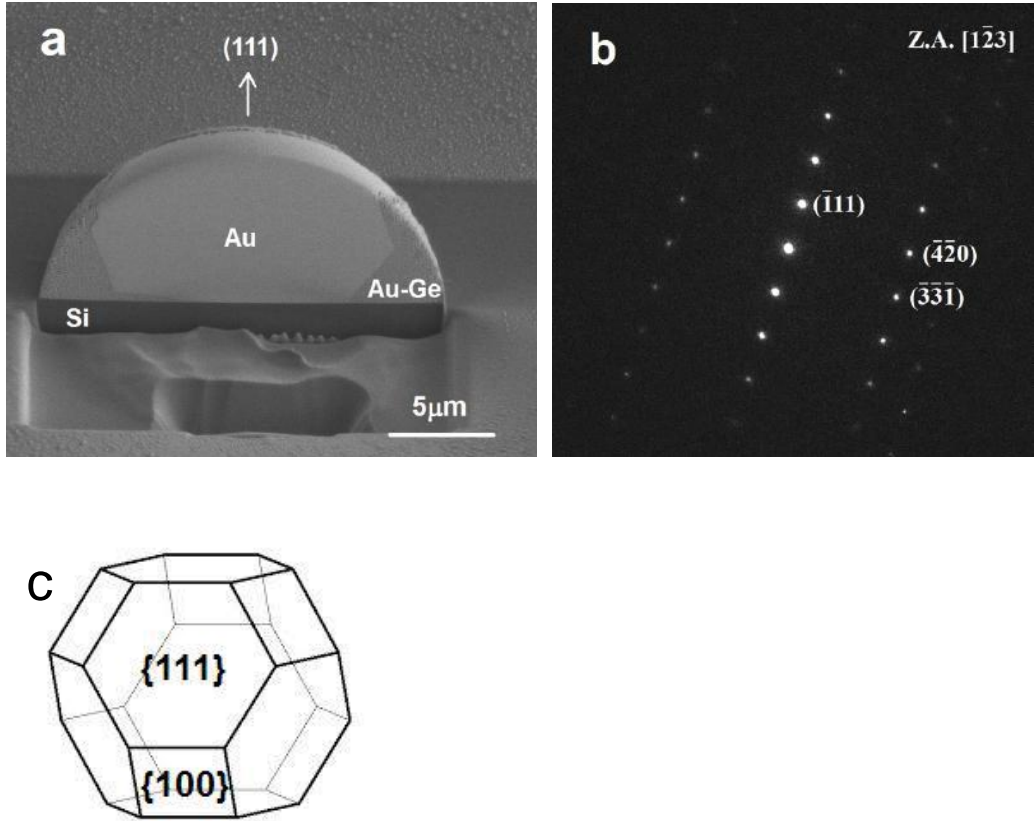

**Figure 2.** (a) HRSEM micrograph of a droplet cross-section obtained by FIB reveals facets inside the droplet but a curved shape on the surface. (b) TEM diffraction obtained from the single crystal area. (c) Truncated octahedron.

As part of our quest to confirm the single crystal nature of the curved gold crystals, we performed Transmission Electron Microscopy (TEM) on thin cross-sections of droplets we had prepared using the FIB (Figure 2(b)). By performing selective area diffraction in the TEM from an area with a diameter of 500nm, we were able to observe a single crystal diffraction pattern that could be fully indexed within the gold structure. Identical electron diffraction patterns were obtained from all parts of the single crystal.

To obtain additional proof that these crystals were single, we used another state-of-the-art characterization technique, sub-micron scanning synchrotron diffractometry (ID13, ESRF, Grenoble, France), on a FIB-sectioned crystal of gold (Figure 3(a)). At each point of the scanned area (1793 scanning points) the same thermal diffuse scattering pattern of a bulk single crystal was observed.[31] We also verified this by averaging all the 1793 diffraction images from the different locations of the crystal and found that this integrated image remained identical to any one of the individual diffraction images (Figure 3(c)-(d)). For quantitative comparison of both diffractions presented in Figure 3(c)-(d), the scattering intensity was plotted for two different azimuthal angles as a function of radial distance (Supporting Information Figure S3). Therefore, it could be seen that the bright diffraction spots in both Figure 3 (c) and (d) generated peaks in the same azimuthal angle and same radial distance (d-spacing). Furthermore, a single crystal Kossel line pattern emerging from Bragg-type diffraction of gold X-ray fluorescence radiation generated by the primary beam was clearly visible in the averaged pattern.[32] An indication of this Kossel pattern, however noisy, can even be seen on the individual patterns (on high quality displays). The lines appear very pronounced when zooming in to a high pass filtered version of the averaged image (Figure 3(e)). These two latter experiments provided conclusive proof that the curved gold crystals were indeed perfect single crystals.

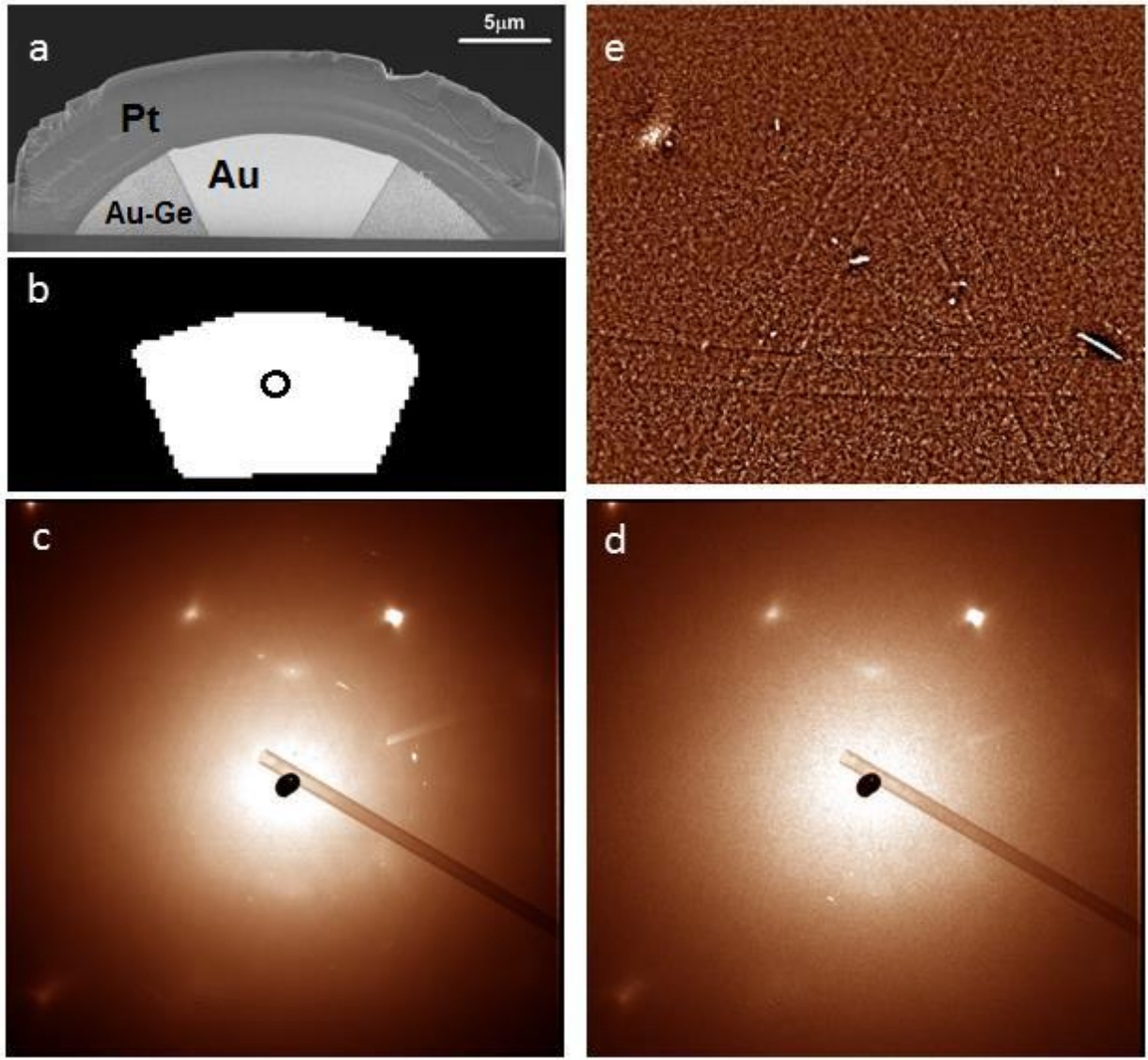

**Figure 3.** (a) Cross-section of a droplet prepared utilizing the FIB which was used for the high-resolution X-ray scan. The droplet was covered with Pt in order to protect it during the ion beam etching. (b) A binary mask derived from mapping of average intensity by simple thresholding in each diffraction pattern, representing X-ray fluorescent background of the scanned gold crystal. (c) Average diffraction pattern of all corrected images of the area in (b) (1793 diffractions). Nano-beam Kossel lines can be observed due to the Au-Fluorescence. (d) Corrected (spatial distortion and flat field) diffraction pattern of scan point which was taken from the center of the single crystal, marked with a circle on (b), for reference. (e) High pass filtered version of the average image enhancing the Kossel-lines.

In addition to the demonstration that it is possible to grow curved single crystals of gold, we wanted to show that the shape of the single crystals can be controlled. This is an extremely important feature for any potential application of such curved crystals. Having shown that the shape of the single crystals is determined by the shape of the confining droplet, we now aimed to control the droplet shape. This can be done by changing the contact angle and the droplet curvature, which is affected by the surface tension. The parameter with the greatest influence on the shape of the droplets was found to be the oxygen partial pressure in the annealing environment. The size of the drop did not influence the droplet shape, as the droplet was much smaller than capillary size (2.6 mm for liquid gold). In samples that were vacuum-pumped ($10^{-3}$ torr) prior to the flow of inert $N_2$ (99.999%) the contact angle was in the range of 90°-110°, whereas for samples prepared under an ambient flow of $N_2$ with no pre-pumping, the contact angle was in the range of 20°-45° (Figure 4). According to the Young-Laplace equation the change in the contact angle is affected by the surface tension of the droplet, the surface energy of the substrate, and the interfacial energy between them. Our results showed that the alteration in oxygen partial pressure changed the contact angle and the surface tension of the droplet.

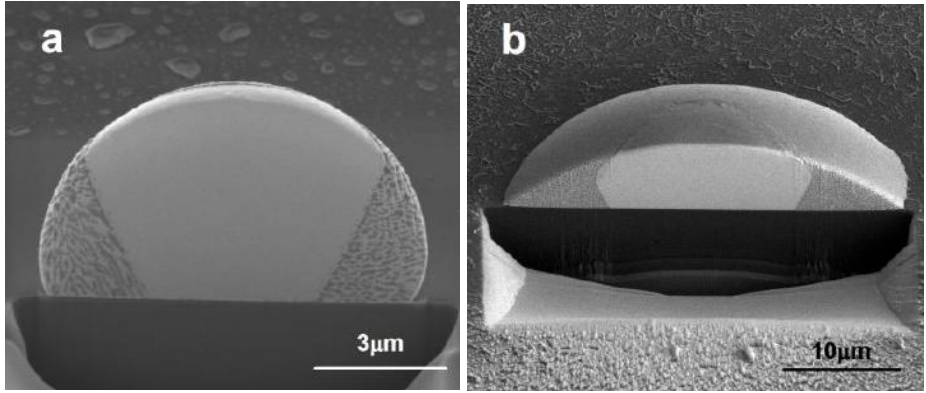

**Figure 4.** The effect of annealing conditions on the droplet contact angle. (a) Eutectic sample annealed with pump-purging prior to annealing. (b) Eutectic sample annealed without pump-purging.

In producing such crystals of various materials, however, several guidance rules or conditions need to be met. These are discussed below.

The concept of growing curved single crystals, as presented above, *i.e.*, by crystallization from a melt within a micro-droplet, is a simple one. However, in translating this concept technically into a generalized phenomenon, a number of guiding principles must apply. Firstly, there must be no interaction (such as diffusion) between the thin film and the supporting substrate, as this would not only lead to a change in the composition of the film but would also enhance wetting (thus obviously preventing dewetting). In the specific system presented here (Au-Ge on Si), a eutectic reaction occurs at the relatively low temperature of 363°C [16] and would instantly lead to rapid diffusion of gold into silicon, thereby instantly shifting the eutectic Au-Ge composition from its original one. Because we have previously found the native Si oxide to be insufficient as a barrier to diffusion, a thick (~100 nm) oxide layer was thermally grown and served as an effective diffusion barrier.[33] Evaporation of gold on the $SiO_2$ was followed by germanium evaporation. This sequence prevented direct contact between Ge and SiO$_2$, which would have led to wetting rather than dewetting.[34, 35]

The second rule is that the composition of the thin films should be chosen such that the ensuing liquid phase causes the crystallization process to be rapid and large single crystals to be formed from the melt (fast diffusion). In our system we utilize the eutectic phase diagram, which shows that in this way the crystallization can be achieved by crossing the solidus line either vertically, by slow cooling, or horizontally, by slow removal of one of the eutectic components. In the Au-Ge system we can either start from a hypoeutectic composition and cross the solidus line vertically, or start from a eutectic composition and cross the solidus line horizontally by annealing in an environment that contains oxygen. It turns out that by this latter route germanium oxidizes finally to GeO, which easily sublimes and leads to GeO evaporation.[36] The result of this process is the precipitation of gold.

The third governing rule is that the material of the thin films should not strongly oxidize before the film dewets, and should melt within a temperature range that will allow dewetting to occur prior to diffusion intermixing with the substrate. In the system used here the eutectic temperature of the Au-Ge is relatively low. Because the Ge thin film is the upper one, its surface harbors a native Ge oxide; however, this oxide desorbs from the surface during the annealing, leaving only Ge on the surface,[36] and therefore does not interfere with the dewetting process.

Note that this method for the growth of curved crystals will work only when all three of the above guidelines are followed. Having shown not only that curved single crystals can be obtained but also that their curvature can be controlled, we wanted to further show that this is not indeed an isolated phenomenon but a general one. We successfully demonstrated this by growing Ge single crystals in the Au-Ge system using a hypereutectic concentration (obtained by changing the initial thickness of the evaporated thin films (120 nm Au, 107 nm Ge) to obtain a hypereutectic concentration of the melt (40at% Ge, Supporting Information Figure S4(a)).

Yet another system we tested was the Ag-Ge thin films, evaporated on the oxidized Si substrates (Ag 40 nm, Ge 212 nm, 20at% Ag), followed by annealing at 850°C for 10 min and then cooling to room temperature at the rate of 200°Cmin$^{-1}$ in a forming gas environment (7.5% H2 + N2, 99.99%). Ge single crystals were observed inside the droplets, confined by the shape of the droplets (Supporting Information Figure S4(b)).

To understand the energetic considerations that allow for growth of the curved single crystals, we have developed an energetic model. Surface energy can be estimated by a number of broken or altered interatomic bonds. The energy required to generate two surfaces of area 2A is the following:

$$E = A n_s (z_b - z_s) E^{ss}, \tag{1}$$

where $E^{ss}$ is the bond energy between the pair of nearest atoms, $z_b$ and $z_s$ are the numbers of nearest neighbors for bulk and surface atoms, and $n_s$ is the surface density of atoms. The number $(z_b - z_s)$ is the number of broken bonds. The surface tension ($\gamma_{sv}$) of the solid/vacuum surface is given by:

$$\gamma_{sv} = \frac{E}{2A} = \frac{1}{2} n_s (z_b - z_s) E^{ss}. \tag{2}$$

The surface tension of the solid/liquid surface can be evaluated in a similar way as

$$\gamma_{sl} = \frac{E}{2A} = \frac{1}{2} n_s (z_b - z_s)(E^{ss} - E^{sl}), \tag{3}$$

where $E^{sl}$ is the bonding energy of a surface atom with neighboring atoms in liquid. Here we assume that the number of nearest neighbors in the liquid and in the solid is the same. This will be true if the liquid is the melt of a solid or a liquid solution based on the same element. From (2) and (3) we obtain:

$$E^{sl} = E^{ss}\left(1 - \frac{\gamma_{sl}}{\gamma_{sv}}\right). \tag{4}$$

Let us compare the change in the Gibbs free energy for two cases: 1) an additional atom adheres to a surface of solid gold under the liquid and remains under the liquid; 2) an additional atom adheres to a surface of solid gold under the liquid but is brought out of the liquid. In the first case the additional atom may increase the number of (s-l) bonds to a value $n_{sl}$. If the atom adheres to the ledge or step, this number will be smaller than in the case of ideal surface. The adherence will cause the following change in the Gibbs free energy:

$$\Delta g_1 = \Delta \mu_{ls} + \Delta E_1, \tag{5}$$

where $\Delta \mu_{ls}$ is the difference of chemical potentials of a gold atom in a bulk solid and liquid, $\Delta E_1 = n_{sl}(E^{ss} - E^{sl})$ is the change in surface energy related to the adhered atom.

In the second case, the Gibbs free energy change will be the following:

$$\Delta g_2 = \Delta \mu_{ls} + \Delta E_2 \tag{6}$$

where $\Delta E_2 = n_{ss}E^{ss} - n'_{sl}(E^{ss} - E^{sl})$ is the corresponding change in the surface energy, $n_{ss}$ is the number of new broken (s-s) bonds generated by the additional atom and $n'_{sl}$ is the number of "closed" (s-l) bonds. The difference of these two energies is given by:

$$\Delta g_2 - \Delta g_1 = \Delta E_2 - \Delta E_1 = (n_{ss} - n'_{sl} - n_{sl})E^{ss} + (n'_{sl} + n_{sl})E^{sl}. \tag{7}$$

In the FCC gold lattice $z_b=12$. At the ideal (111) surface the atom has $z_s=9$ nearest neighbors. If an additional Au atom adheres to this surface, it may have 3 nearest bonds in the case of the ideal surface, 6 bonds if it adheres to the ledge, and 9 bonds – to the step. In the limit case, when the energy change $\Delta E_1$ is maximal and the $\Delta E_2$ is minimal: $n_{sl}=6$, $n_{ss}=3$, $n'_{sl}=3$, the energy difference (7) is minimal. Combined with relation (4) it is given by:

$$\Delta g_2 - \Delta g_1 = 9E^{sl} - 6E^{ss} = 3E^{sl}\left(\frac{\gamma_{sv} - 3\gamma_{sl}}{\gamma_{sv} - \gamma_{sl}}\right). \tag{8}$$

As can be seen, this difference is positive for $\gamma_{sl} < \frac{1}{3}\gamma_{sv}$. Applying the values of gold surface energies, $\gamma_{sv} = 1.283 \frac{J}{m^2}, \gamma_{sl} = 0.187 \frac{J}{m^2}$,[37, 38] one can conclude that adherence of the gold atoms from the liquid to the solid/liquid interface is more favorable than adherence to the facets entering from the liquid to a vapor.

**Conclusions**

Using relatively simple methods such as dewetting from thin films, and exploiting thermodynamic phenomena such as crystallization from a eutectic melt, we have demonstrated a method for creating a single curved crystal that grows in the confined space of its own melt, replicating the shape of the droplet. To achieve this phenomenon a number of factors must coexist. The substrate must be inert relative to the thin films, dewetting of the thin film melt must be achieved so as to create droplets on the surface, crystallization from a melt must be possible in the relevant environment, and the materials must not react (for example *via* oxidation) with the environment.

**Methods**

*Sample preparation*. SiO$_2$ (100 nm) was grown on (001) Si wafers by thermal oxidation at 1100°C. Gold and germanium (99.999% pure, Sigma-Aldrich) were evaporated onto the SiO$_2$ substrate in an e-beam-equipped AircoTemescal FC-1800 evaporating system under a high vacuum of 10$^{-7}$ Torr, at room temperature, yielding a normal deposition rate of 8 Ås$^{-1}$. Thermal annealing of Au-Ge samples was performed at 550°C in a nitrogen environment (99.999%) for 10 min to 3 h in a laboratory tube oven with gas flow, and in a DHS 1100 Domed Hot Stage (Anton Paar) connected to a rotation pump.

*Sample characterization*. For surface imaging by Scanning Electron Microscopy (SEM) we used an FEI ESEM Quanta 200 (combining EDS and EBSD) and a Zeiss Ultra plus HR-SEM.

Cross-section micrographs were obtained using a Strata 400 STEM dual-beam FIB field emission scanning electron microscope. Transmission electron microscopy diffraction was obtained utilizing a Titan 80-300 KeV FEG-S/TEM. The single crystal was characterized with the micro-diffraction technique at the ID13 Microfocus Beamline of the European Synchrotron Research Facility (ESRF), Grenoble, France. One branch of this beamline is specialized on scanning nano-beam diffraction experiments. Synchrotron light focused by means of Si-based compound refractive lenses in crossed geometry, so-called nano-focusing lenses, at a wavelength of 0.832 Å was utilized for the experiment. The beam size was 170 X 150 nm$^2$ (Full width of half maximum of the beam intensity) delivering a flux of $10^{10}$ photons/sec in the focal spot. The sample position was controlled *via* piezo stages with 20 nm resolution. Two-dimensional maps of diffraction patterns have been recorded using a FRELON CCD detector with 2048x2048 pixels of 50 μm in size.

AUTHOR INFORMATION

**Corresponding Author**

*Boaz Pokroy

E-mail: bpokroy@technion.ac.il

Department of Materials Science and Engineering, Technion Israel Institute of Technology, 32000 Haifa, Israel.

*Conflict of Interest:* The authors declare no competing financial interest.

*Acknowledgment.* The research leading to these results has received funding from the European Research Council under the European Union's Seventh Framework Program (FP/2007–2013)/ERC Grant Agreement n° [336077]. We also acknowledge the European COST Action TD0903. Thin films were fabricated at the Micro-Nano Fabrication Unit (MNFU) at the Technion, Haifa. We are grateful to Dr. Tzipi Cohen-Haims, Dr. Alex Berner and Michael Kalina for their help with preparing samples and operating the electron microscopes. We are also grateful to Dr. Yaron Kaufman for his help with TEM. The nano-focusing lenses have been kindly provided by the group of Prof. Dr. Christin Schroer from TU-Dresden, Germany.

*Supporting Information Available:* Figure S1, EDS spectrum of the gold single crystal. Figure S2, EBSD results for the gold single crystal. Figure S3, scattering intensity of thermal diffuse scattering pattern plotted for two different azimuthal angles as a function of radial distance. Figure S4, Ge single crystal in the Au-Ge system and Ge single crystal in the Ag-Ge system. Figure S5, Cross section of the thin film before annealing.

This material is available free of charge *via* the Internet at http://pubs.acs.org.

**ToC figure**

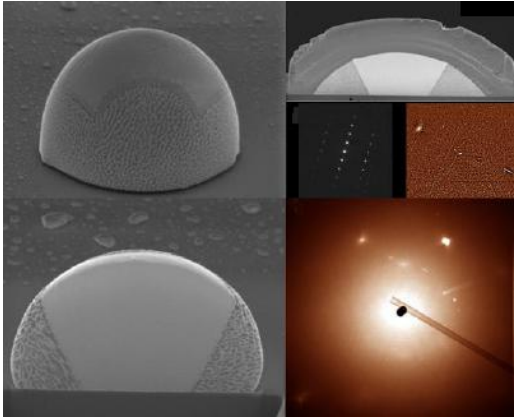